\documentclass{article}

\usepackage{amsmath}
\usepackage{enumerate}
\usepackage{graphicx}
\usepackage{hyperref}

\newcommand{\cor}[1]{#1}

\begin{document}

\title{Brownian Thermometry Beyond Equilibrium}

\date{}

\author{Daniel Gei\ss \thanks{Institute for Theoretical Physics, University of Leipzig, Germany} \thanks{Max Planck Institute for Mathematics in the Sciences, Leipzig, Germany} \and Klaus Kroy\footnotemark[1] }

\maketitle

\section{Abstract}

Since Albert Einstein's seminal 1905-paper on Brownian motion, the temperature of fluids and gases of known viscosity can be deduced 
from observations of the fluctuations of small suspended probe particles. We summarize recent generalizations of this standard technique of Brownian thermometry to situations involving spatially heterogeneous temperature fields and other non-equilibrium conditions in the solvent medium. The notion of effective temperatures is reviewed and its scope critically assessed. Our emphasis is on practically relevant real-world applications, for which effective temperatures have been explicitly computed and experimentally confirmed. We also elucidate the relation to the more general concept of (effective) temperature spectra and their measurement by Brownian thermospectrometry. Finally, we highlight the conceptual importance of non-equilibrium thermometry for active and biological matter, such as microswimmer suspensions or biological cells, \cor{which often  play the role of non-thermal (``active'') heat baths for embedded Brownian degrees of freedom.}

\section{Introduction}
Brownian motion is the jiggling of microscopic particles suspended in a liquid or gaseous medium. Named after Robert Brown, who meticulously investigated the vexing incessant erratic motion of dissolved pollen fragments through a microscope in the 1820s, its physical origin remained obscure until the early 20th century. It was Albert Einstein in his ``annus mirabilis'', who eventually ended a long time of confusion and heated debates with his paper ``On the movement of small particles suspended in a stationary liquid demanded by the molecular-kinetic theory of heat'' \cite{Einstein}.  By a clever combination of simple physical arguments, he established a straight link between the mesoscopic phenomenon of Brownian motion and the molecular composition of all matter. In the following years, the theoretical ideas were experimentally confirmed by Jean Baptiste Perrin (Later awarded with the Nobel Prize in Physics in 1926) and used to prove the existence of atoms and molecules, which is the key to a mechanistic explanation of the ubiquitous Brownian fluctuations and the notions of heat and temperature. The mathematical concept of Brownian motion spread to a wide range of applications, exceeding the classical realm of physics to subjects in chemistry, biology and even economics \cite{Kroy2}.  What is sometimes lost in these interdisciplinary applications and abstractions is the origin of Brownian motion from the thermal motion of the molecules of a physical medium. It is of course precisely this connection that is exploited to deduce the solvent temperature from quantitative measurements of the Brownian fluctuations of suspended probe particles. It is the very basis for the notion of Brownian thermometry, the topic of this contribution. For a detailed discussion of alternative thermometry methods at the nanoscale see for example \cite{Carlos}. Moreover, a theoretical discussion of the minimal requirements of a thermometer to  determine  the  temperature  of a system as well as a mechanical model for a thermometer in non standard cases is given in \cite{baldovin2017thermometers}.
\par
\begin{figure}[h]
	\centering
	\includegraphics[width=0.8\textwidth]{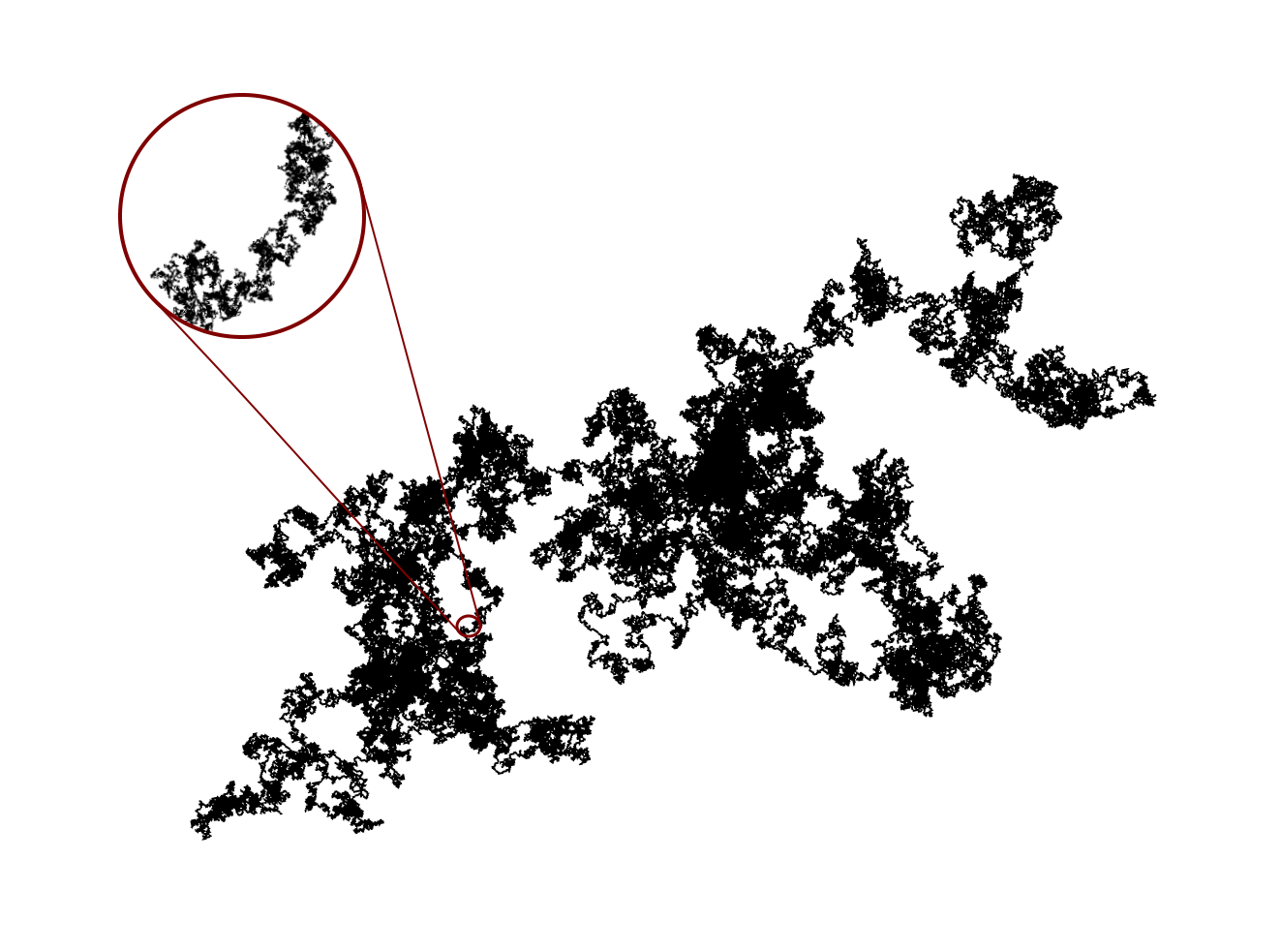}
	\caption{\textbf{The fractal paths of Brownian motion.} Brownian particles suspended in a solvent medium undergo an erratic thermal motion tracing out a rugged path called a random walk (here illustrated in two dimensions). Their trajectories are statistically self-similar, with enlarged subsections qualitatively resembling the whole. In Brownian thermometry, one deduces the temperature of the solvent from observations of these trajectories.}
	\label{fig:frog}
\end{figure}
In equilibrium, Brownian thermometry may just seem as a somewhat arduous way of measuring the solvent temperature, redundant with other, more convenient techniques. Yet, one obvious advantage is that it employs a strongly localized probe that can be inserted, say, inside a living cell. But why should one want to extend it to situations with non-equilibrium baths, as announced in the title? First and foremost, one may argue that local temperature measurements become most interesting when the molecular temperature of a system is not homogeneous in space and time, i.e., if the system is not isothermal. Secondly, Brownian particles have recently been used to realize microscopic heat engines in contact with non-equilibrium heat baths \cite{Krishnamurthy,Zakine,Martinez,Holubec}. The ability to define and control the pertinent effective bath temperatures is crucial for assessing the efficiency of such non-equilibrium heat engines. Finally, non-equilibrium Brownian thermometers afford us with paradigms for how to approach non-equilibrium fluctuations in more complex real-world applications, e.g., in active and living matter \cite{leshouches,bechinger2016active,marchetti2013hydrodynamics}, which is not only macroscopically perturbed from equilibrium but strongly driven on a molecular or mesoscopic scale.  
\par
\par
The remainder is structured as follows. In the second section we summarize the basics of Brownian thermometry in equilibrium. Therefor, we recapitulate the main arguments put forward by Einstein in his work from 1905, elucidating the origin and significance of the Einstein relation. Subsequently, we introduce its generalization, the so-called fluctuation-dissipation-theorem (FDT), to further clarify the deep connection between thermal fluctuations and the notion of temperature with the energy loss due to friction. Section \ref{Section 3} focusses on Brownian motion in non-equilibrium situations. Avoiding technical details, we discuss why the definition of temperature beyond equilibrium is not unique and introduce the much-debated notion of effective temperatures. Our special focus is on the ``hot Brownian motion'' of a colloidal particle that is heated above the ambient temperature. The heating gives rise to a persistent ``housekeeping'' heat flow into the non-isothermal solvent and to enhanced non-equilibrium Brownian fluctuations of the colloid that can be characterized by a set of effective temperatures. From these exactly computable effective temperatures an explicit generalized Einstein relation (GER) and a generalized FDT ensue. \cor{Hot Brownian motion is an experimentally and theoretically fully accessible, highly symmetric minimum extension of the central paradigm of equilibrium stochastic thermodynamics to strongly non-isothermal conditions. As such it provides a paradigm for many ramifications of the same theme that are currently the subject of intense research in the field of so-called active matter which we adress in Sec. 4.3. Of particular interest in this respect are asymmetric hot Brownian particles, since they are natural examples of self-propelled microswimmers} \cite{Kroy4}
\cor{. The study of their non-equilibrium stochastic and directed autonomous motion has established a tractable model for more complex artificial nano-machines and living matter with potential applications in technology and medicine. We suggest that one important benefit of these studies is to generate elements of a metric to quantify deviations from equilibrium} We close with a brief summary and outlook.
\section{Equilibrium Brownian Thermometry} \label{Section 1}
\subsection{Einstein's Pioneering Work from 1905} \label{Section 1.1}
Einstein considered a small particle in an isothermal solvent \cite{Einstein}. Assuming that this particle is denser than the surrounding solvent, one observes a persistent average downward drift due to gravity. However, the fluctuations around this overall trend prevent the particle from permanently settling to the ground. Their thermodynamic effect can be described as an osmotic pressure gradient from regions of high particle concentration to regions of low concentration. In thermal equilibrium, the gravitational and osmotic forces are perfectly balanced, giving rise to the Brownian barometer equation for the particle concentration $c$ as function of height $h$
\begin{equation}
	c(h) = c(0) \mathrm{e}^{-U(h)/k_\mathrm{B}T}, \label{Brownian barometer equ.}
\end{equation}
where $U(h)$ is the (buoyancy-corrected)  gravitational potential above ground.
Concomitant with this balance of forces one expects a balance of particle fluxes in equilibrium. According to Stokes' law, the downward force $-\nabla U$ leads to the average downward drift flux $-c\nabla U/\zeta$ with $\zeta$ denoting the friction of the particles with the solvent. The overall effect of the Brownian fluctuations can be subsumed into a counterbalancing systematic diffusion flux $-D\nabla c$, as stated by Fick's law. Combining the \textit{detailed balance}  of fluxes with equation \eqref{Brownian barometer equ.} yields the Einstein relation
\begin{equation}
	D = \frac{k_\mathrm{B} T}{\zeta}, \label{Einstein relation}
\end{equation} 
which was actually first found by Sutherland \cite{Sutherland}. As it links the strength of thermal fluctuations, encoded in the diffusion coefficient $D$, to the strength of dissipation in transport, encoded in the friction coefficient $\zeta$, it can be regarded as the mother of all fluctuation-dissipation relations. The remarkable feature of this relation is that it ties together two at first sight unrelated transport coefficients of the mesoscopic world and exposes their common origin from the thermal notion of atoms via the ``thermal energy quantum'' $k_\mathrm{B}T$.
\par
From a more pragmatic point of view \eqref{Einstein relation} highlights the possibility to use a Brownian particle as a thermometer. Assuming the friction coefficient $\zeta$ of the particle in the solvent to be known, the temperature $T$ can be inferred from a measurement of the diffusion constant $D$. Alternatively, equation \eqref{Einstein relation} together with an explicit formula $\zeta(\eta)$ for the friction can also be used to employ the Brownian particle as a passive micro-rheometer for the solvent viscosity $\eta$, if the temperature is known \cite{Furst}. 
\subsection{The Fluctuation-Dissipation-Theorem} \label{Section 1.2}
Before addressing more advanced problems, far from equilibrium, we want to gather some intuition for the physics near equilibrium. We therefor recapitulate two slight reformulations and generalizations of \eqref{Einstein relation}, namely, Onsager regression and the fluctuation-dissipation relation (FDT). Based on Einstein's work on thermal fluctuations \cite{Einstein,Einstein2}, Onsager \cite{Onsager} formulated his remarkable \textit{regression principle}: the relaxation of macroscopic non-equilibrium disturbances is governed by the same laws as the regression of spontaneous microscopic fluctuations in an equilibrium system. In other words, the decay of  forced (weak) deviations from equilibrium  cannot be distinguished from the one of spontaneous fluctuations.
This general principle is a cornerstone of many modern time-dependent applications in statistical and thermal physics and one of the main achievements that earned Onsager the  Nobel Prize in Chemistry in 1961. It can itself be understood as a direct consequence of the FDT\footnote{Onsager formulated his principle in 1931, two decades before general proofs of the FDT were published, and one may speculate that Onsager was also aware of the FDT  \cite{Chandler}.}. 
\par
To establish the FDT, imagine a macroscopic object embedded in a fluid, which is set into motion by some external means. Obviously, its motion is not in equilibrium. But, due to friction, which arises from disordered collisions with molecules of the solvent (that also acts as a heat bath), the motion will slow down.  The kinetic energy of the object is thereby converted to heat --- in other words, it is dissipated. 
This motivates the fundamental assumption of linear response: if a macroscopic variable $X$ is perturbed from equilibrium by an amount $\Delta X$, the system will experience a restoring force $F\propto - \Delta X$ proportional to the perturbation $\Delta X$ itself. Although a microscopic justification is tricky \cite{vanKampen}, its statistical-mechanics derivation is straightforward and provides the foundation of \textit{linear response theory} \cite{Chandler}. It is the starting point for an extension of equilibrium statistical mechanics to situations close to equilibrium and hence provides a good description for many practically interesting phenomena. 
\par
The example illustrates how thermal molecular motion turns mechanical energy into heat, and thereby drives the system dissipatively towards equilibration. But, in addition to dissipation, we also observe fluctuation effects as the other side of the same coin, so to say. It is most noticeable for microscopic or mesoscopic particles that they never entirely come to rest.  And even if a colloidal particle of mass $m$ in a solvent could  initially be held still, it would instantly start moving, its speed and direction permanently  changing in a random fashion, as a consequence of random collisions with bath molecules. In equilibrium, the effects of Brownian fluctuations and dissipation balance each other on average, such that the solvent affords the average kinetic energy $3k_\mathrm{B}T/2$ to the colloid.  
\par
This is the essential physical content of the FDT. To formalize it, consider, as a tutorial toy model, the position $x$ of a single colloid obeying the Langevin equation
\begin{equation}
	m\ddot{x} = -\gamma \dot{x} + R.  \label{Langevin equ}
\end{equation}
The first term on the right is immediately recognized as a type of linear response and corresponds to friction, i.e. the systematic collective restoring force exerted by the solvent onto the colloid to slow it down deterministically. The term $R$ comprises the remaining ``unsystematic'' part of the colloid-solvent interactions, due to random contributions of the collisions with the solvent molecules. On sufficiently long time scales, on which the friction effect is well described by the Markovian term -$\gamma \dot{x}$, $R$ can be modeled as infinitely rapidly fluctuating. For consistency with standard equilibrium statistical mechanics, its statistical properties are then prescribed as follows:
\begin{enumerate}[(i)]
	\setlength{\parskip}{2pt}
	\setlength{\itemsep}{0pt plus 1pt}	
	\item it has vanishing average, i.e. 
	\begin{equation}
	\langle R(t) \rangle=0,
	\end{equation}
	\item it is independent of the previous history of the colloid velocity $\dot x(t)$ and
	
	\item not correlated with itself, so that it can be idealized as 
	\begin{equation}\label{eq:FDT}
		\langle R(t_1)R(t_2) \rangle= C\delta(t_1-t_2)
	\end{equation}
	with the Dirac $\delta$-function and some constant ``noise strength''  $C$ (to be determined).
\end{enumerate}
Only if $C$ is chosen appropriately, the average $\langle \dots \rangle$, to be taken with respect to the distribution of the realizations of the stochastic force $R(t)$, is consistent with the usual Gibbs-Boltzmann averaging procedures of equilibrium statistical mechanics. To find this appropriate value of $C$, one can consider the solution of \eqref{Langevin equ}, which takes the form
\begin{equation}
	\dot x(t) = \dot x(0) \mathrm{e}^{-\frac{\gamma}{m} t} + \frac{1}{m}\int_0^t \mathrm{d}t'\ \mathrm{e}^{-\frac{\gamma}{m} (t-t')} R(t')  .
\end{equation}
The first term on the right remembers (for a time $m/\gamma$) the initial velocity $\dot x(0)$, which decays due to friction. The second term describes a sum over all velocity fluctuations caused by molecular collisions. They are likewise damped (or ``filtered'') by friction, as required by Onsager's regression hypothesis.
\par
At late times $t\rightarrow \infty$, we expect the system to equilibrate. Consistency check with equipartition for the average kinetic energy, namely
\begin{equation}
	\langle \dot x^2 \rangle = \frac{k_\mathrm{B}T}{m},
\end{equation}
yields $C=2\gamma k_\mathrm{B} T$, turning (\ref{eq:FDT}) into the so-called \textit{fluctuation-dissipation-theorem}  
\begin{equation}
	2\gamma k_\mathrm{B} T = \int_{-\infty}^{+\infty} \mathrm{d}s\ \langle R(0) R(s) \rangle  \qquad \quad \text{(FDT)}.   \label{FDT}
\end{equation}
\par
This result is one of the most important tools in (classical) statistical physics, and slight generalizations apply to quantum mechanics as well.  
\cor{It represents a general relationship between the linear response of a given system to an external disturbance and the internal fluctuation of the system in absence of the disturbance.} 
\par
Already Einstein noticed, that his results could readily be generalized to electric circuits (a connection later made more quantitative by Johnson and Nyquist \cite{Nyquist,Johnson}). Since then, many similar relations for all kinds of transport coefficients were derived. All these relations introduced new possibilities to measure and better understand the complicated behavior of strongly interacting many-particle systems. Incidentally, they also provide an explanation and remedy for the breakdown of the deterministic formalism of thermodynamics at critical points, where susceptibilities diverge and fluctuations become long ranged. And they can be employed to precisely quantify the associated anomalies, such as the dynamic slowdown \cite{Domb}.
\section{Non-Equilibrium Brownian Thermometry} \label{Section 3}
\subsection{Temperature Beyond Equilibrium} \label{Section 3.1}
Although the precise meaning and most suitable definition of temperature in small equilibrium systems is still sometimes debated in the literature \cite{Vazquez,Dunkel,Frenkel}, equilibrium statistical mechanics is now generally quite well understood. The same cannot be said about systems out of equilibrium. They are still a very active field of research, posing a wide variety of problems. Interesting topics of recent research are systems with slow relaxation (e.g. aging in molecular glasses), strongly driven systems  (e.g. granular matter or strongly perturbed colloidal systems), and active matter (comprising, e.g., living organisms and artificial microswimmers) \cite{leshouches,Gompper,bechinger2016active,marchetti2013hydrodynamics}. 
\par
A key feature of equilibrium statistical mechanics is that it provides a means to accurately describe macroscopic systems in terms of a small number of thermodynamic variables (e.g. temperature, pressure).  Far from equilibrium an analogous general framework is not available, and the very existence of key concepts of thermodynamics such as pressure or temperature is \emph{a priori} questionable \cite{Solon}. Nevertheless, such systems often can to large extent be described in terms of so-called effective parameters which act as replacements for their corresponding equilibrium counterparts.  In the following, we focus on the aspect of temperature, but it should be noted that progress has also been made for other thermodynamic quantities (e.g.\ for the pressure \cite{steffenoni}).
\par
The formal definition of temperature (for a detailed discussion see for example \cite{puglisi2017temperature,miller1952concept}) essentially rests on the fundamental laws of thermodynamics. The zeroth law asserts the transitivity of mutual thermalization and thereby provides the conceptual basis for conventional temperature measurements by physical contact or other weak interactions.  However, such empirical temperature, operationally defined solely on this basis, does not yet provide a universal reference for what is hot or cold. The second law of thermodynamics provides such a reference and gives temperature a universal succinct meaning. It first establishes the meaning of ``hotter'' and ``colder'' by stating the direction of the flow of heat from hotter to colder bodies, and, in consequence, even a universal absolute temperature scale can be defined. In addition, the second law allows for the definition of entropy. Via the Gibbs relation
\begin{equation}
	T = \frac{ \partial U}{\partial S} \bigg|_{V,N},   \label{Gibbs relation}
\end{equation}
the key concepts of temperature and entropy become connected.
\par
In systems that are not globally equilibrated but can in some sense be linked to an equilibrium state, the local equilibrium hypothesis is applicable. Most often, one can assume that the system can be divided into subsystems or compartments that are large enough to contain a huge number of microscopic degrees of freedom but small enough to be considered homogeneous and unperturbed. Assigning the corresponding equilibrium values on the small scales, one finds that thermodynamic quantities such as temperature and pressure can be defined as fields $T(\vec r,t)$ and $p(\vec r,t)$, varying in space and/or time. Even if the system is driven very far from global equilibrium, far reaching statements about its fluctuations can then be made as long as the system is globally isothermal \cite{Seifert}.
\par
For systems that somehow defy the local equilibrium hypothesis, the explicit treatment becomes much more challenging. The zeroth law of thermodynamics, which is an important element in the conventional definition of temperature, is then generally not applicable. Generalized notions of temperature therefore often need to rely on the (first and) second law, alone. Independent definitions of temperature may then yield different values for the temperature of the same system, since energy does not have to be equally distributed between diverse degrees of freedom \cite{Falasco6}. To illustrate the point, imagine a system composed of matter and electromagnetic radiation, and we assume to have two different thermometers, one which is sensitive to the matter content and another to radiation. If the system is in thermal equilibrium both thermometers will measure the same temperature. Otherwise, the results may differ. In consequence, one finds a possibly large number of different definitions of temperature and realizations of thermometers, some typical examples being \textit{dynamical temperatures} \cite{Rugh,Rugh2}, \textit{contact temperatures} \cite{Muschik,Muschik2}, \textit{effective
temperatures} \cite{Cugliandolo} and \textit{generalized temperatures} \cite{Jepps,Rickzayen}. We content ourselves with a brief overview and refer the reader to the review of Casas-V\'{a}zquez and Jou \cite{Vazquez} for a more comprehensive discussion.
\par
In absence of a zeroth law, there are various approaches how to generalize the notion of temperature. One relies on entropy and entropy fluxes from which temperature is deduced  for example by the Gibbs relation \eqref{Gibbs relation}. This seems to be a safe starting point, but it requires a good control over the pertinent notion of entropy, which may be difficult to come by in practice. \cor{Even if it is known (which is usually not the case), the absolute entropy may not be helpful, for example because it is constant in a non-equilibrium steady state (NESS), which renders the Gibbs relation} \eqref{Gibbs relation} \cor{less useful.} Alternatively, one may introduce an effective temperature $T_\mathrm{eff}$ by extrapolating macroscopic equations of state (e.g.\ for the internal energy or magnetization). 
This establishes a mapping to an equilibrium situation by replacing the temperature $T$ of the equilibrium equations of state by $T_\mathrm{eff}$. In consequence, the so defined effective temperature depends on some reference temperature and on some perturbing fields (electric field, velocity field, temperature gradient, etc.). The problem arising in this context is that it is in general questionable whether equations of state retain their form in non-equilibrium situations (and indeed the example of hot Brownian motion discussed below provides an exactly understood counter example). Instead of relying on the deterministic thermodynamic relations to generalize temperature, one might rather resort to fluctuations. If one can establish a generalized FDT for some pertinent mesosopic variables one can assign some effective temperatures to quantify their fluctuations relative to their (non-equilibrium) response.  The strategy was pursued (at least) as early as in 1989 by Hohenberg and Shraiman \cite{Hohenberg} in the context of weak-turbulence, who formally defined an effective temperature by the departure from the FDT. But a deeper understanding of its thermodynamic status (if any) remained elusive, since it is not easily extended to a Gibbs relation \eqref{Gibbs relation}, which would relate it to entropy. Later, the concept was adopted for granular matter by S. F. Edwards \cite{Edwards,Edwards2}. More recent applications to glasses, advocated e.g.\ by Cugliandolo and coworkers, evolved along exactly solvable schematic models \cite{Cugliandolo3,Cugliandolo4}. They provided a better understanding of the scope of effective temperatures far from equilibrium and put the notion on a firmer ground. 
\par
Fluctuation dissipation violations are generally observed in non-equilibrium systems. A problem that may therefore easily arise in this approach is that diverse non-equilibrium contributions to the FDT get lumped into one effective temperature, regardless of their possibly diverse origins (e.g. from non-linear generalizations of the FDT \cite{Stratonovich}). 
In this context, G. S. Agarwal \cite{Agarwal}, in his pioneering work, discussed systems driven out of equilibrium and corresponding to the steady-state solution of the Fokker-Planck equation. He found possible generalized FDTs that are not easily reconciled with the equilibrium formulation by a mere effective temperature.
Also more recent works studying the FDT in sheared colloidal suspensions \cite{Fuchs} or other strongly driven soft matter systems \cite{Seifert2} suggest an additive (not multiplicative) extension of the FDT which is not straightforwardly interpreted in terms of $T_\mathrm{eff}$. To emphasize the fundamental difference of such additive corrections to the FDT, compared to those that one would naturally want to capture by an effective temperature, some researchers have given the name ``frenesy'' to the additional term appearing in the linear response \cite{baiesi}.
It has been suggested that the effective temperature might still be a sensible definition when the system evolves slowly, irrespectively of whether it relaxes or is in a steady state \cite{Cugliandolo5}. Furthermore, Cugliandolo et al. \cite{Cugliandolo2} were able to show that in situations with a small energy flow, such an effective temperature can indeed play the role of a thermodynamic temperature, in the sense that it identifies the direction of heat flow (i.e.\ heat flows from regions with larger to those with smaller values of $T_{\text{eff}}$). It thereby provides a criterion for thermalization reminiscent of the zeroth law, meaning that systems in mutual equilibrium will have the same value of $T_{\text{eff}}$ \cite{gnoli2014nonequilibrium}. \cor{However, this familiar property is not strictly necessary and unavoidable, as in the case of conventional temperatures characterizing some local equilibrium. And it is} indeed not shared by the exactly computable and experimentally measurable effective temperatures discussed in the next section, which arise in situations with large heat fluxes. \cor{The crux is that these effective temperatures characterize the fluctuations of Brownian degrees of freedom that couple differently and independently to their non-equilibrium environment. They do not excite mutual heat fluxes nor mutually equilibrate. While such unruly behavior may seem a nuisance, at first, it can also be turned into an advantage. Temperatures confined to selective degrees of freedom (e.g. to the center of mass of a Brownian particle but not to its molecular vibrations) allow extraordinarily high ``thermal'' gradients to be maintained without burning the machinery, with formidable consequences for the efficiency} \cite{Krishnamurthy}. \cor{Similarly unintuitive effects arise upon cooling selective particle degrees of freedom} \cite{Arita,Millen,Kroy5}.
\subsection{A Perfect Paradigm: Hot Brownian Motion} \label{Section 3.2}
To speak of a temperature in the traditional sense, it must be measurable by a thermometer. In contrast, \cor{without a zeroth law,} effective temperatures and their appropriate thermometers are not independent of the design of the bath and the device assessing the heat. So, often the heat-consuming device itself will be the best-adapted (if not the only) thermometer to measure a pertinent effective temperature. But even this strategy may not yield unique results. A Brownian tracer particle in a non-equilibrium bath provides an instructive example. It can be used as its own (optimally adapted) thermometer. But notice that one may either assign a temperature to its diffusivity via the generalized Einstein relation or, alternatively, to its kinetic energy via the equipartition rule. As it turns out, these procedures do not generally yield the same result, far from equilibrium. An early numerical comparison of apparently incompatible notions of temperature in the context of non-isothermal Brownian motion was performed in \cite{Barrat2}. At first sight, this study might seem to stand against the generalization of Brownian thermometry to situations out of equilibrium. However, as we discuss next, this is actually not the case.
\par
To avoid some less relevant technical difficulties, we focus on the highly symmetric case of \textit{hot Brownian motion} (HBM) \cite{Rings,Falasco,Falasco2} of a spherical colloid that acts  both as a tracer and as a persistent heat source. The heat emanating from a hot Brownian particle spreads quickly through the fluid, establishing a comoving stationary temperature gradient (in the sense of a spatially varying local equilibrium) around the relatively slowly moving particle. Colloidal particles can also persistently be cooled relative to the surrounding solvent, although this is experimentally much more challenging \cite{Roder}. But theoretically, it just amounts to a sign change, so that such \textit{cold Brownian motion} is implicitly included in our discussion. For an explicit mathematical description of HBM one would like to repeat Einstein's derivation, but one cannot resort to such powerful equilibrium concepts as equipartition and Stokes' law to leapfrog over the explicit evaluation of the colloid-solvent interactions. Instead, the equations describing the non-equilibrium dynamic fluctuations of the non-isothermal solvent have to be coarse-grained, explicitly. As suggested by common sense, the resulting equations of motion for the particle look similar to the equations \eqref{Langevin equ} for the equilibrium case and have a similarly universal character (although somewhat less than in equilibrium). In essence, a hot particle surrounded by a molecular temperature field $T(r)$ in the fluid will simply diffuse faster compared to an isothermal Brownian particle in the same fluid with constant temperature $T=T(r\rightarrow\infty)$. The only tricky question being by how much exactly. The role of the effective temperature is then to define a mapping of the motion of the hot particle to an equilibrium Brownian motion at temperature $T_\mathrm{eff}=T_\mathrm{HBM}$. An important point to notice is that $T_\mathrm{HBM}\neq T(r=0)$ which means that a hot Brownian particle is not equilibrated locally.
\par
\begin{figure}[h]
	\centering
	\includegraphics[width=0.5\textwidth]{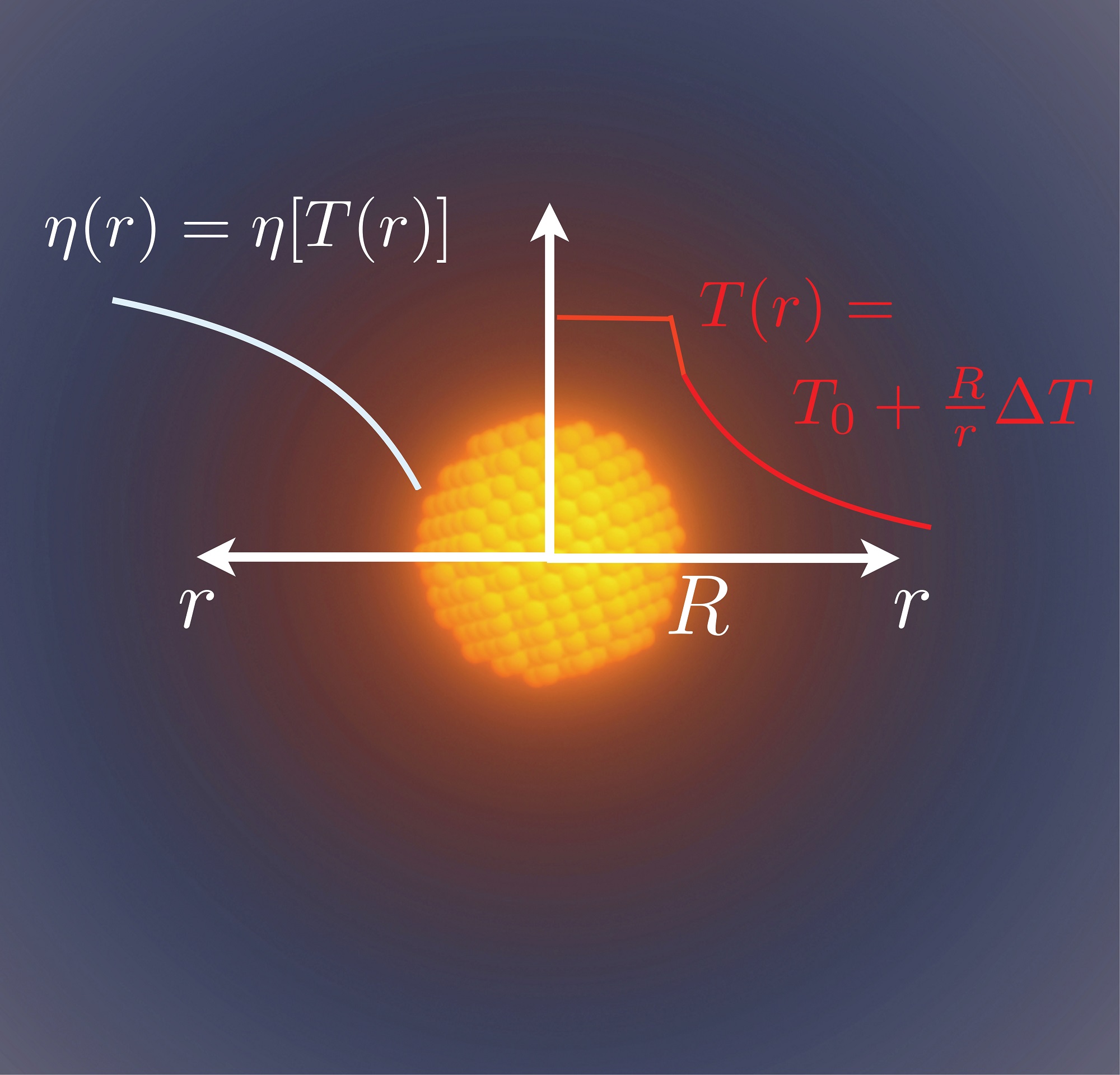}
	\caption{\textbf{Hot Brownian motion of a spherical colloid.} The molecular temperature and viscosity field fields, $T(r)$ and
	$\eta(r)$, around a uniformly heated Brownian particle of radius $R$. Such particles perform a non-equilibrium Brownian motion that appears to be equilibrated at an effective temperature $T_\text{HBM}=T_0+5\Delta T/12$, intermediate between the surface temperature $T_0+\Delta T$ and the ambient temperature $T_0$. Closer inspection reveals a multitude of analytically computable effective temperatures, as described in the text. (Adapted from Ref. \cite{Kroy})}
\end{figure}
As alluded to above, practical challenges of the sketched program lie in the explicit calculation of the effective temperature, and some other effective transport coefficients. For example, in addition to temperature, also the viscosity of the solvent will vary in the vicinity of the hot particle. Therefore, not only the pertinent effective temperature of the thermal fluctuations that drive the particle's motion is \emph{a priori} unknown, but also the effective viscosity and friction coefficient that attenuate it. In the Markov limit, corresponding to the low time resolution already considered in deriving the Langevin equation \eqref{Langevin equ}, memory effects in the solvent dynamics can be ignored \cite{Li, Franosch}. In this case, one can indeed show that the non-equilibrium effects can be encapsulated in effective parameters, such as an effective reduced friction coefficient  $\zeta_\mathrm{HBM}$ and an effective Brownian temperature $T_\mathrm{HBM}$.
These parameters have to be calculated from the underlying non-isothermal fluctuating hydrodynamic theory \cite{Falasco2, Sengers} by a systematic coarse graining procedure that integrates out the long-ranged hydrodynamical interactions between the particle and the fluid. This task can be performed analytically for a hot (or cold) Brownian sphere.  One finds that rotation and translation, exciting different solvent flow fields, have their own effective temperatures \cite{Chakraborty,Rings2}. Far from equilibrium, temperature is thus seen to  loose its universal character and to become more of an interaction parameter between the particle and its solvent, akin to the equilibrium friction coefficients $\zeta$, say, which also differ between rotation and translation.  In general, the effective parameters are complicated functions of the molecular temperature and viscosity fields $T(\vec r)$, $\eta(\vec r)$, and the solvent velocity field excited by the particle motion. Nevertheless, they can be calculated in good precision in many situations of practical interest. In a second step, the effective diffusion coefficient $D^\alpha_\mathrm{HBM}$ can then be derived to establish generalized Einstein relations \cite{Falasco,Chakraborty} 
\begin{equation}
	D^\alpha_\mathrm{HBM} = \frac{k_\mathrm{B}T^\alpha_\mathrm{HBM}}{\zeta^\alpha_\mathrm{HBM}},   \label{eff Diffusion Coef}
\end{equation}
where $\alpha\in\{\mathrm{t,r} \}$ labels the degrees of freedom (translations and rotation).
\par
In summary, the Sutherland-Einstein relation \eqref{Einstein relation} can be generalized from ordinary equilibrium Brownian motion to nonisothermal (``hot'') Brownian motion by the simple replacement of the thermodynamic parameters by their effective counterparts. Thus, we infer that a hot Brownian particle can still be used as a thermometer or rheometer, but with the important caveat what is measured is an effective temperature or viscosity, respectively. Since different values for $T^\alpha_\mathrm{HBM}$ pertain to different degrees of freedom $\alpha$, one can however even draw some conclusions as to the spatial structure of the ``conventional'' molecular temperature field $T(\boldsymbol{r})$ from such measurements, although this inverse problem clearly does not possess a unique solution, in general. \cor{We also note that the above notion of effective temperature is generalizable to other fields of application, like electrical engineering, nanophotonics} \cite{Skolnik,Milligan}\cor{, and possibly to active fluctuations in living matter} \cite{Turlier,Kroy3}\cor{. We also reiterate that an important reason to define, measure, and compute these effective temperatures is that they are bona-fide temperatures in the sense of the second law for the considered degrees of freedom. They therefore enter all kinds of thermodynamic considerations such as efficiencies etc.} \cite{Krishnamurthy}. \cor{Another incentive, exemplified below, is that they may be the key to discovering and formalizing hidden but robust symmetries of non-equilibrium Brownian motion. But before we get to this, there is one more remark to make.}
\par
\cor{Namely, as it turns out, one can push the analysis of effective temperatures one major step further, and include memory effects. For a brief explanation, we turn again to the model of HBM which we have so far discussed in the Markov limit. In the technically more involved general case, the Brownian dynamics is not Markovian, and} each effective temperature can be shown to be replaced by a frequency dependent \textit{temperature spectrum} $\mathcal{T}^{\alpha}(\omega)$ \cite{Falasco}. The effective temperatures $T^{\alpha}_\mathrm{HBM}$ governing the Markov limit are recovered in the long-time or low-frequency limit $\mathcal{T}^{\alpha}(\omega\to0)\to T^{\alpha}_\mathrm{HBM}$. But for fast motion or high frequencies they smoothly cross over to new effective temperatures that characterize the translational and rotational \emph{kinetic} degrees of freedom of the particle (as opposed to the positional ones). These are the temperatures that appear if one attempts to generalize the equilibrium equipartition theorem for the particle velocities, which are apparently (effectively) equilibrated at these various new temperatures. But again, due to the lack of a zeroth law, the kinetic energy is actually not equally distributed between the different degrees of freedom, nor do they mutually equilibrate. For a hot particle in a cool solvent, the kinetic degrees of freedom are always hotter than the spatial ones, and for example rotation is hotter than translation \cite{Rings2}. 
\par 
Considering a weakly damped hot Brownian particle trapped in a harmonic potential --- a situation of high practical interest e.g.\ for nanoparticles controlled by optical tweezers \cite{Juan,Li2} --- then leads to the fancy notion of ``\textit{Brownian thermospectrometry}'' \cite{Falasco3}. The characterizing observable of the setup can be chosen to be the particle position $x(t)$ and the confinement force is $m\omega_0^2x(t)$. If the particle is weakly damped by the solvent, \cor{essentially only the resonant mode $k_\mathrm{B}\mathcal{T}(\omega_0)$ of the thermal spectrum} is absorbed by  the particle. The appropriate weak-coupling limit can be realized for large particle-to-fluid mass density ratios (e.g. Brownian particle suspended in gas \cite{Li,Millen,Li2}). Then the mean energy of the particle obeys the familiar equipartition theorem (with a reduced mass $M$ accounting for the effect of the solvent inertia on an accelerated suspended particle)
\begin{equation}
	m\omega_0^2 \langle x^2 \rangle = M \langle \dot x^2 \rangle = k_\mathrm{B}\mathcal{T}(\omega_0).
\end{equation}
\cor{By varying the trap stiffness $m\omega_0^2$, the oscillator represented by the trapped particle can be tuned to various frequencies $\omega_0$ to sample the spectrum $\mathcal{T}(\omega)$ for translation (and similarly for rotation), so that one can interpret the set-up as a thermospectrometer. By combining this information from several degrees of freedom one can increasingly constrain the mentioned ``reverse problem'' of inferring the molecular temperature field $T(r)$.}
\par 
So far, a number of predictions of the theory of hot Brownian motion, as sketched above, could be validated with great precision by experiments \cite{Rings,Selmke,Radunz,Ruijgrok,Falasco4} and non-equilibrium molecular dynamics simulations \cite{Chakraborty,Falasco4}. Conversely, after the accuracy of the theory has been established, it can help to develop practically useful innovative experimental applications, such as photothermal microscopy/spectroscopy setups, which employ nano-particles suspended in water as dynamic tracers  \cite{Selmke,Radunz,Berciaud}. Gold particles can for example be heated by a green laser beam that is well absorbed, and the mirage generated by the emanating heat in the solvent can be detected by another (red) laser beam. A useful feature of such a setup is that it allows for a highly sensitive back-ground-free lock-in detection, so that it can be recommended as a more robust, sensitive and easily manageable replacement for similar fluorescent tracer techniques. Also the experimental analysis of rotational hot Brownian dynamics has found its use in thermometry applications with rotational nanomotors realized by dissolved gold nanorods driven by polarized light \cite{Andren}. 
\subsection{Towards Complexity: Active and Biological Matter}\label{Section 3.3}
\cor{We have argued above that hot Brownian motion provides an important test bed for the notion of effective temperature and non-equilibrium thermometry. It is both exactly solvable and precisely measurable. Most real-world phenomena are not. This naturally raises the question whether this perfect paradigm can guide us through more complex terrain. }
\par
Life is an active process far from equilibrium --- in equilibrium you are dead. But an unambiguous  distinction between activity and thermal noise is not always easy, in particular, if one needs to detect stationary active processes that do not change in time. Systems in thermodynamic equilibrium are not only stationary in time, which means that rates of transitions between different microstates are overall balanced. They obey the much stricter symmetry that their transitions are balanced pairwise. It was L. Boltzmann in 1872 \cite{Boltzmann} who first noticed this property of detailed balance and thereby provided the foundation of equilibrium statistical mechanics. In living matter, stationary stochastic processes may appear to be thermally driven, at first sight, but can still turn out to be more complex, upon further inspection.  Since the days of Robert Brown, it has thus remained a central question in the field how one can distinguish active from equilibrium thermal fluctuations and how activity or liveliness can be quantified. In fact, active fluctuations have been shown to be superimposed on thermal fluctuations by a careful analysis of FDT violations in living matter, down to the scale of (relatively) simple red blood cells \cite{Turlier}.
\par
One particularly intriguing feature of living matter is the locomotion and transport of cells, molecules, organelles, and microorganisms, which facilitates the organization and assembly of complex organisms, chemotaxis, sexual evolution and signaling, or the formation of colonies and swarms. \cor{All these phenomena produce environments that act on embedded Brownian tracer particles as ``active'' (a priori non-thermal) baths.} Nature has invented a large number of designs for transport and self-propulsion. Moreover, natural microswimmers, such as bacteria, algae and sperm, which usually appear in large numbers and exhibit well organized behavior, forming large-scale patterns such as networks, vortices or swarms, which may even coordinate to fulfill macroscopic tasks. While very interesting in their own right, they can sometimes be difficult to control and study. This has spurred the development of laboratory models for small-scale locomotion, so-called artificial microswimmers. These artificial biomimetic toy models may potentially help to make experimental studies more reproducible and to exert better experimental and theoretical control over the complex emergent behavior found in living matter.  Again, artificial nano- or microswimmers are of interest in their own right, promising a diverse range of innovative technical and medical applications.  A recent collection of mini-reviews on the topics of biological and artificial swimmers as well as collective motion can be found in \cite{Gompper}. 
\par	
To refer to the common traits of such living and biomimetic systems, the notion of ``active  matter'' has been coined for this important class of non-equilibrium many-body systems that are not driven by some external forces but by ``intrinsic'' energy-consuming mechanisms. The active mechanisms break detailed balance on a microsopic or mesoscopic scale.
But, in principle, it can reemerge at larger scales \cite{ben2011effective}. For example, for the mentioned microswimmer solutions, breaking of detailed balance is obvious on a mesoscale, where the active transport is visible, but much less so on a macroscopic scale, where they may resemble ordinary colloidal suspensions, in many ways \cite{steffenoni}.
Therefore, to test for the breaking of detailed balance in active matter, the dynamics should be examined at various scales and from multiple perspectives. A detailed discussion that compares several different approaches to resolve this issue can be found in the review of Gnesotto et.\ al \cite{Broedersz}. Their emphasis is on the (non-invasive) detection of circular currents in configuration phase space, which have been shown to provide a reliable route \cite{Zia,Broedersz2}. They are forbidden in equilibrium by the principle of detailed balance and therefore provide an unambiguous footprint of life.  
\par
As already mentioned for the red blood cells, FDT violations may serve as an additional valuable method to check for activity. In particular, if the FDT violations have a multiplicative structure,  as for example for the red blood cells \cite{Turlier},
activity or liveliness may be measured in terms of effective temperatures. For a large class of such ``simple'' (or ``non-frenetic'') NESS states, the departure from equilibrium may thus be related and quantitatively compared via their effective temperatures \cite{ben2011effective}. In other words, effective temperatures may provide a quantitative metric that deals with the ``intrinsic'' activity of animate and inanimate matter on an equal footing.  The dimensionless ratio $T_\mathrm{eff}/T$ --- which becomes unity in equilibrium --- then affords us with an absolute universal metric to measure deviations from equilibrium, in a similar way as we use the absolute Kelvin scale to quantify the distance from the limit of a  purely mechanical or quantum-mechanical system ``at zero temperature''. To obtain an explicit value for the effective temperature, experimentally, one has to measure a linear response function, i.e., one has to perturb the variables of interest. To perform such controlled perturbations can be technically demanding and may sometimes even lead to undesired effects. Moreover, it may not always be possible to reliably compute the effective temperatures from first principles. Yet, neither is an entirely hopeless task, as the above discussion of hot Brownian motion has demonstrated. In the final paragraph we want to show how this can be extended to the case of microswimmers, and how it can be combined with yet another general method of quantifying dissipative fluxes and the distance from equilibrium, namely so-called fluctuation theorems.    
\subsection{A Universal Hierarchical Metrik for NESS}
\begin{figure}
	\centering
	{\includegraphics[width=0.5\textwidth]{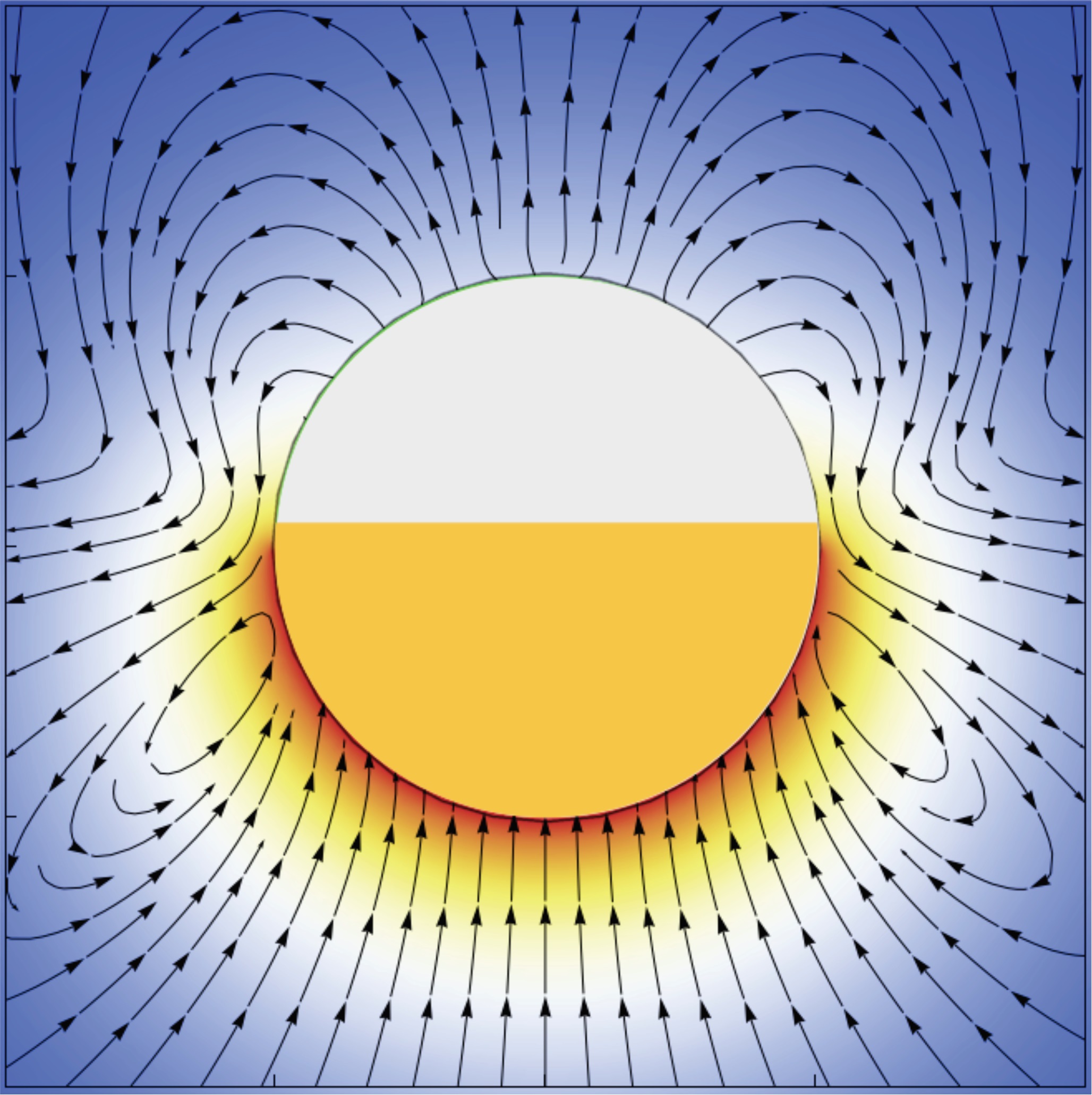}}
	\caption{ 
		\textbf{Hot Brownian microswimmer.} Schematic of a Janus sphere half covered with gold. Heating by a leaser beam creates an asymmetric temperature field in the surrounding solvent, which excites thermoosmotic currents that give rise to a directed self-phoretic motion of the Janus particle along its symmetry axis. (Adapted from Ref. \cite{Kroy,Kroy4,Bickel} )
		} \label{fig:janus}
\end{figure}
\par
Hot Brownian particles are easily turned into microswimmers by breaking their geometric or thermal symmetry. The archetypal design is illustrated in Fig.~\ref{fig:janus}, which shows a so-called Janus sphere, half covered with gold \cite{Jiang}. Under non-isothermal conditions, such Janus spheres, and, in fact, any thermally asymmetric particles, possess a directed motion. Any such heated or non-isothermal non-symmetric particle can thus be classified as a \emph{hot Brownian swimmer}. The asymmetric temperature field in its surrounding solvent excites a boundary-layer flow along the particle surface \cite{Bickel,Kroy4}, leading to a directed motion along the symmetry axis. Due to the rotational hot Brownian motion, the direction of the symmetry axis will permanently change and ultimately randomize the overall motion again. In result, one finds hot Brownian motion at short times $t\ll D^\mathrm{t}_\mathrm{HBM}/v_\mathrm{p}^2$, ``ballistic'' swimming motion at intermediate times $D^\mathrm{t}_\mathrm{HBM}/v_\mathrm{p}^2\ll t \ll 1/D^\mathrm{r}_\mathrm{HBM}$, and diffusive motion $\langle r^2(t)\rangle=2v_\mathrm{p}^2t/D^\mathrm{r}_\mathrm{HBM}$ at late times $t \gg 1/D^\mathrm{r}_\mathrm{HBM}$, with the rotational diffusion coefficient defined in \eqref{eff Diffusion Coef}. \cor{Note that, on a large scale, the locally persistent but ultimately stochastic motion gives rise to yet another type of effective Brownian temperature, }$T_\mathrm{eff}=\frac{v_\mathrm{p}^2 \zeta}{k_\mathrm{B} D^\mathrm{r}_\mathrm{HBM}}$,\cor{ if one applies the Einstein relation, }Eq.~\eqref{Einstein relation} to $\langle r^2(t)\rangle$.\cor{This effective temperature does indeed enter the barometer equation for a suspension of active particles} \cite{palacci2010sedimentation}. \cor{A fluid of many microswimmmers can thus be seen as a prototypical active bath that transmits non-equilibrium ``active'' fluctuations to an embedded passive tracer. The latter can then itself behave similar to an active particle.} 
\par
\cor{While some reader may at this point be bewildered by the seemingly endless proliferation of effective temperatures and transport coefficients, which do indeed need to be handled with some care, we want to emphasize again that they are not created artificially or by invoking new principles and concepts. Quite to the contrary, they arise from the multifaceted dynamic features that become possible far from equilibrium, precisely  because of the lack of an important equilibrium principle, namely the zeroth law.}
\par

\cor{We now want to explain another virtue of the various effective thermodynamic notions. Namely, the effective quantities can help to discover and formulate some robust hidden symmetries that hold even far from equilibrium, such as the so-called fluctuation-theorems and thermodynamic uncertainty relations. }\cor{They can be understood as generalizations of the FDT and the second law of thermodynamics and constrain arbitrarily large driven or spontaneous fluctuations.} They come in several different (but closely related) formulations and hold for a wide range of non-equilibrium systems \cite{Seifert,Evans2,shankar2018hidden,falasco2019unifying,barato2015thermodynamic}. For the concrete case of our hot Brownian swimmer, the broken time-reversal invariance due to the active (energy-consuming) swimming motion is precisely quantified by the part $\mathcal{S}$ of the entropy production associated with the active (directed) motion. The active motion itself can be considered a dissipative particle flux that turns work supplied by its propulsive engine into heat at a rate $\dot Q$. Importantly, this dissipation rate is responsible for the mentioned entropy production via $\dot{\mathcal{S}} = \dot Q/T_\mathrm{HBM}$, with the appropriate effective temperature for translational hot Brownian motion  \cite{Falasco4}. With these definitions, the fluctuation theorem for the hot Brownian swimmer takes the form of an exponential relation
\begin{equation}\label{eq:FT}
	P(\mathcal{S}) = P(-\mathcal{S}) \mathrm{e}^{\mathcal{S}/k_\mathrm{B}}
\end{equation}
between the probabilities $P(\mathcal{S})$ and $P(-\mathcal{S})$ for forward and backward moves of the swimmer, corresponding to positive ($\mathcal{S}$) and negative ($-\mathcal{S}$) entropy production, respectively. 
For the hot Brownian swimmer, this fully explicit formulation of the fluctuation theorem could not only be analytically derived, it was also verified with high accuracy by experiment and massively parallel computer simulations \cite{Falasco4}. 
\cor{Moreover, it was found that the hot Brownian swimmer satisfies and even saturates the universal dissipation bound set by the thermodynamic uncertainty relation. This is significant, since it asserts once more that the hot swimmer's motion, while possibly extremely strongly driven away far from equilibrium in one sense (e.g. in terms of heat or particle flux), is still within the linear response regime of an effective steady state with detailed balance. And that this holds irrespective of its detailed technical realizations, which actually differ substantially between the numerical simulation (a Lennard-Jones cluster with symmetric wetting properties in a Lennard-Jones fluid) and the experiments (a gold-polystyrene Janus particle in water). In particular, it is not limited to a Maxwellian velocity distribution} \cite{shankar2018hidden,falasco2019unifying,barato2015thermodynamic}.

\begin{figure}[h]
	\centering
	{\includegraphics[width=1.0\textwidth]{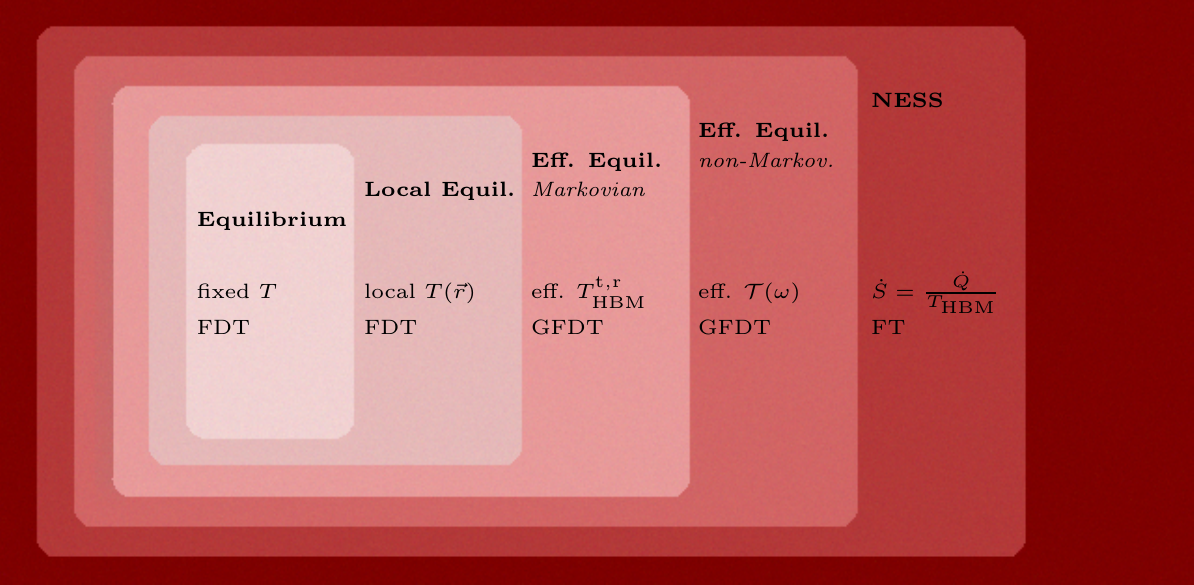}}
	\caption{ 
		\textbf{Hierarchy of non-equilibrium steady-states (NESS).} For a variety of systems very far from equilibrium, violations of the fluctuation-dissipation theorem (FDT) are not additive but multiplicative. They can be mended by introducing effective temperatures, yielding a generalized FDT (GFDT). Such effective temperatures may be well-defined and explicitly computable, even if the zeroth law of thermodynamics is not applicable. Yet, the NESS class for which this has so far been achieved is by no means exhaustive. A metric hierarchy involving effective temperatures along with exact fluctuation theorems for dissipative currents is proposed to quantitatively measure the distance from equilibrium for more ``frenetic'' systems, for which the deviations from equilibrium cannot be captured by effective temperatures alone. The notation refers to the actual experimental system described in the text, a so-called hot Brownian swimmer, for which an exact analytical theory and high precision data from experiments and massively parallel non-equilibrium molecular-dynamics simulations are available.
		} \label{fig:scheme}
\end{figure}

The results of this analysis thus underscore once more the usefulness of the effective temperatures.  Finally, notice that the exponential of the entropy production in the fluctuation theorem (\ref{eq:FT}) quantifies the distance from equilibrium in a hierarchical way:   firstly, via the dimensionless distance $T_\mathrm{HBM}/T$ from isothermal Brownian motion in terms of the effective temperature, and secondly via the entropy production associated with the dissipative active ``swimming'' flux. It has an explicit representation in terms of the dissipation to a \emph{virtual heat bath}\footnote{Recall that the solvent is non-isothermal so that there is no actual (only a virtual) equilibrium bath, here.} at the effective temperature $T_\mathrm{HBM}$.  This is exactly the temperature measured by the ``device'' itself, namely by the Janus swimmer (if interpreted as a hot Brownian particle in the comoving frame). 
The emerging metrical hierarchy for measurements of the distance from equilibrium is, for the discussed example, schematically illustrated in Fig.~\ref{fig:scheme}. The various effective temperatures of hot Brownian motion are crucial ingredients of this scheme.
\section{Conclusion and Outlook}
In this work, we have reviewed recent progress in Brownian thermometry, i.e., the inference of the temperature of a fluid from observations of the motion of suspended Brownian particles. We have emphasized the deep connection between fluctuations and dissipation, which give rise to a special symmetry of all types of thermal equilibrium dynamics, encoded in the fluctuation-dissipation theorem (FDT) and the principle of detailed balance. Our main interest was in situations far from equilibrium, where temperature is \emph{a priori} no longer well-defined. In this regime, the zeroth law of thermodynamics, which is conventionally used to introduce the notions of temperature and thermometers, does not necessarily hold anymore. Differently designed thermometers then cease to measure the same quantity. As a consequence, there are various compatible and incompatible proposals of how to generalize and assess temperature(s) under such strong non-equilibrium conditions. We have emphasized recent progress towards the rigorous definition of effective temperatures, based on an explicitly computable generalization of the fluctuations-dissipation theorem for a class of paradigmatic non-equilibrium steady states (NESS). Our pet example, hot Brownian motion, describes the thermal motion of a particle that is persistently maintained at an elevated (molecular) temperature relative to the surrounding fluid (e.g.\ by irradiation). The exact analytical coarse-graining of this non-isothermal system affords us with a precise and well-defined notion of explicitly computable effective temperatures that were confirmed with high precision, in experiments and computer simulations. On slow time scales, in the so-called Markov limit, these results provide a non-trivial mapping onto conventional equilibrium Brownian motion, so that the well-tested procedures and protocols developed for equilibrium Brownian thermometry can straightforwardly be taken over to the non-equilibrium case. By calculating a frequency dependent effective temperature, it is even possible to venture beyond this convenient practice and include short-time memory effects. In future high-speed setups, this (so far) purely theoretical development could thus give rise to useful practical applications of a more elaborate thermometric technique that is best characterized as Brownian thermospectrometry in non-isothermal solvents.  
Currently, there is great interest in exploring possible generalizations of these precise notions of non-equilibrium thermometry to more complex situations involving active and biological matter, where the FDT and detailed balance are broken by energy-consuming microscale and mesoscale processes. As a practical example that illustrates  how the various concepts introduced may fruitfully be applied to a paradigmatic active-matter system, we discussed a so-called hot Brownian swimmer. The exactly computed effective temperatures were integrated into a hierarchical metric scheme that can fully quantify the distance from equilibrium of its experimentally measured motion. \cor{This establishes them as crucial ingredients of any survey of the vast and largely uncharted landscape of non-equilibrium many-body dynamics.}
\par	
\section{Acknowledgements}
We acknowledge funding by the International Max Planck Research Schools (IMPRS), the Deutsche Forschungsgemeinschaft (DFG) via the priority program SPP 1726 ``microswimmers'' and by the Humboldt foundation. We thank Gianmaria Falasco for helpful discussions.
\bibliographystyle{plain}
\bibliography{bibfile}
\end{document}